\pdfoutput=1
\documentclass[runningheads]{llncs}
\usepackage{graphicx}
\usepackage{balance}  
\usepackage{xspace,multirow,multicol,colortbl,xcolor,tabularx}
\usepackage[ruled, vlined, linesnumbered]{algorithm2e}
\SetKwComment{Comment}{$\triangleright$\ }{}
\usepackage{subfig}
\usepackage[font=bf]{caption}
\usepackage{comment}
\usepackage{enumitem}
\usepackage{amssymb}
\usepackage{amsmath}
\usepackage{flushend}
\usepackage{tikz}
\usepackage{pgfplots}
\pgfplotsset{compat=1.16}
\usetikzlibrary{patterns}
\usepackage{cite}

\makeatletter
\newcommand*\bigcdot{\mathpalette\bigcdot@{.5}}
\newcommand*\bigcdot@[2]{\mathbin{\vcenter{\hbox{\scalebox{#2}{$\m@th#1\bullet$}}}}}
\makeatother

\def\done{\hspace*{\fill} {$\square$}}
\def\header{\vspace{1mm} \noindent}

\newcommand{\AM}{\mathbf{A}\xspace}
\newcommand{\DM}{\mathbf{D}\xspace}

\newcommand{\PM}{\mathbf{P}\xspace}

\newcommand{\MM}{\mathbf{M}\xspace}

\newcommand{\HM}{\mathbf{H}\xspace}
\newcommand{\QM}{\mathbf{Q}\xspace}

\newcommand{\SM}{\mathbf{S}\xspace}
\newcommand{\TM}{\mathbf{T}\xspace}
\newcommand{\IM}{\mathbf{I}\xspace}

\newcommand{\ie}{{\it i.e.},\xspace}
\newcommand{\eg}{{\it e.g.},\xspace}

\newcommand{\csmt}{$\mathtt{Co\textrm{-}Simmate}$\xspace}
\newcommand{\fsim}{$\mathtt{F}\textrm{-}\mathtt{CoSim}$\xspace}
\newcommand{\dsim}{$\mathtt{D}\textrm{-}\mathtt{CoSim}$\xspace}
\newcommand{\piter}{$\mathtt{PowerMethod}$\xspace}
\newcommand{\rsim}{$\mathtt{RPCS}$\xspace}

\newcommand{\csmtb}{$\boldsymbol{\mathtt{Co\textrm{-}Simmate}}$\xspace}
\newcommand{\fsimb}{$\boldsymbol{\mathtt{F}\textrm{-}\mathtt{CoSim}}$\xspace}
\newcommand{\piterb}{$\boldsymbol{\mathtt{PowerMethod}}$\xspace}

\newlength\lengtha 

\newenvironment{customlegend}[1][]{%
    \begingroup
    \csname pgfplots@init@cleared@structures\endcsname
    \pgfplotsset{#1}%
}{%
    \csname pgfplots@createlegend\endcsname
    \endgroup
}%

\def\addlegendimage{\csname pgfplots@addlegendimage\endcsname}

\makeatletter
\newcommand\footnoteref[1]{\protected@xdef\@thefnmark{\ref{#1}}\@footnotemark}
\makeatother

\let\oldnl\nl
\newcommand{\nonl}{\renewcommand{\nl}{\let\nl\oldnl}}

\begin{document}
\title{Fast Approximate All Pairwise CoSimRanks via Random Projection}
%
%
\author{Renchi Yang\thanks{Work done while at NUS} \and Xiaokui Xiao}

\authorrunning{Yang and Xiao.}
%
\institute{Hong Kong Baptist University, National University of Singapore\\
\email{renchi@hkbu.edu.hk}, \email{xkxiao@nus.edu.sg}}
\maketitle              
\begin{abstract}
Given a graph $G$ with $n$ nodes, and two nodes $u,v\in G$, the {\em CoSimRank} value $s(u,v)$ quantifies the similarity between $u$ and $v$ based on graph topology. Compared to SimRank, CoSimRank is shown to be more accurate and effective in many real-world applications including  synonym expansion, lexicon extraction, and  entity relatedness in knowledge graphs. The computation of all pairwise CoSimRanks in $G$ is highly expensive and challenging. Existing solutions all focus on devising approximate algorithms for the computation of all pairwise CoSimRanks. To attain a desired absolute accuracy guarantee $\epsilon$, the state-of-the-art approximate algorithm for computing all pairwise CoSimRanks requires $O(n^3\log_2(\ln(\frac{1}{\epsilon})))$ time, which is prohibitively expensive even $\epsilon$ is large. In this paper, we propose \rsim, a fast randomized algorithm for computing all pairwise CoSimRank values. The basic idea of \rsim is to approximate the $n\times n$ matrix multiplications in CoSimRank computation via random projection. Theoretically, \rsim runs in $O(\frac{n^2\ln(n)}{\epsilon^2}\ln(\frac{1}{\epsilon}))$ time and meanwhile ensures an absolute error of at most $\epsilon$ in each CoSimRank value in $G$ with a high probability. Extensive
experiments using six real graphs demonstrate that \rsim is more than up to orders of magnitude faster than the state of the art. In particular, on a million-edge Twitter graph, \rsim answers the $\epsilon$-approximate ($\epsilon=0.1$) all pairwise CoSimRank query within 4 hours, using a single commodity server, while existing solutions fail to terminate within a day.

\keywords{CoSimRank, random projection, approximate algorithm.}
\end{abstract}
\section{Introduction}
Measuring node similarity is a fundamental problem in graph analysis and mining. CoSimRank \cite{rothe2014cosimrank} is recently proposed as a powerful similarity measure for quantifying the similarity between two nodes in a graph based on graph topology. CoSimRank finds numerous applications, such as {\em synonym expansion} and {\em lexicon extraction} in natural language processing, linguistically-informed statistical tool in {\em Cistern} project \cite{cistern}, modeling {\em entity relatedness} in knowledge graphs \cite{ponza2017two,zeng2019measuring}, as well as measuring user similarity in social networks \cite{liao2019second}. 

CoSimRank is closely related to existing similarity measures such as {\em Personalized PageRank} (PPR) \cite{jeh2003scaling,haveliwala2003topic} and {\em SimRank} \cite{jeh2002simrank}. Given a graph $G$ and two nodes $u,v\in G$, the PPR value of node $v$ w.r.t node $u$ is defined as the probability that a random walk starting from node $u$ terminates at node $v$, which measures the strength of connections between two nodes instead of their neighborhood similarity. In contrast, the basic idea of SimRank is that two nodes are considered to be similar if their neighbors are similar. More specifically, the SimRank value of node pair $(u,v)$ considers the first meeting time of random walks from $u$ and $v$. Combining the ideas of PPR and SimRank, the CoSimRank value $s(u,v)$ aggregates all possible meeting times of random walks from $u$ and $v$, resulting in improved effectiveness than PPR and SimRank in many applications, as shown in \cite{rothe2014cosimrank}, but incurring tremendous computational overheads.

Existing solutions \cite{rothe2014cosimrank,yu2015co,yu2018fast} towards CoSimRank computation all focus on all-pairwise CoSimRank queries. Specifically, given an input graph $G$ with $n$ nodes, the all-pairwise CoSimRank query asks for approximate CoSimRank values of all possible node pairs in $G$. \piter \cite{rothe2014cosimrank} directly applies {\em power iterations} \cite{page1999pagerank} to approximate CoSimRank values iteratively, which involves expensive matrix multiplications as well as numerous iterations to ensure $\epsilon$ absolute error guarantee in each CoSimRank value. \fsim \cite{yu2018fast} accelerates CoSimRank computation for given query nodes by first constructing the spanning polytree structures through breadth-first search. When the query set is the node-set of $G$, \fsim has the same time complexity as \piter. \csmt \cite{yu2015co} reduces the number of iterations required in \piter by reusing the intermediate results from previous iterations to speed up the computation in further iterations. In many practical scenarios \cite{rothe2014cosimrank,yu2015co,yu2018fast}, high-precision results are not necessary since a relatively large absolute error guarantee $\epsilon$, \eg $0.1$, in each CoSimRank value is sufficient for our purpose to identify the similar nodes. Unfortunately, all these methods entail $O(n^3)$ time for matrix multiplications in each iteration, to attain the desired accuracy $\epsilon$, even $\epsilon$ is large. Though we can harness the power of computing cluster to compute the CoSimRank values w.r.t each node independently in parallel, each thread still suffers from an $O(n^2)$ time cost, which is prohibitive for large graphs. Thus, the retrieval of all pairwise CoSimRanks remains a highly challenging problem.

Motivated by this, we propose \rsim (short for \underline{R}andom \underline{P}rojection-based \underline{C}o\underline{S}imRank), an efficient algorithm for approximate all pairwise CoSimRank computation. The basic idea is to perform dimensionality reduction on the $n\times n$ random walk matrix of the input graph $G$ to obtain an $n\times d$ matrix ($d\ll n$), such that we can approximate the matrix multiplications in the $d$-dimensional space in an efficient manner, thereby avoiding the $O(n^3)$ time cost in the course of computing CoSimRanks. Using Johnson–Lindenstrauss transformation \cite{johnson1984extensions,achlioptas2003database,matouvsek2008variants} for the dimensionality reduction, \rsim is able to provide an accuracy guarantee $\epsilon$ on each CoSimRank value in terms of its absolute error with a high probability. Additionally, \rsim needs the same number of iterations as in \piter but each iteration consumes $O(\min\{n^2/\epsilon^2,n^3\})$ time, which is favorable and satisfies our requirements when $\epsilon < \frac{1}{\sqrt{n}}$. Extensive
experiments using six real graphs demonstrate that \rsim is more than up to orders of magnitude faster than the state of the art. In particular, on a million-edge Twitter graph, \rsim answers the $\epsilon$-approximate ($\epsilon=0.1$) all pairwise CoSimRank query within 4 hours, using a single commodity server, while existing solutions fail to terminate within a day.

The rest of the paper is organized as follows. Section \ref{sec:back} provides the necessary background for CoSimRank and the formal problem definition. Related work is reviewed in Section \ref{sec:rw}. In Section \ref{sec:algo}, we present our proposed \rsim method and related analysis. Our solution and existing methods are evaluated in Section \ref{sec:exp}. Finally, Section \ref{sec:clc} concludes the paper.
\section{Preliminary}\label{sec:back}

\begin{table*}[!t]
\centering
\renewcommand*{\arraystretch}{1.4}
\begin{small}
\caption{Frequently used notations.}\vspace{2mm}\label{tbl:notation}
\begin{tabular}{|c|l|}
\hline
\bf Notation & \bf Description \\ \hline
$G=(V,E)$ & The input graph $G$ with node set $V$ and edge set $E$. \\ \hline
$n, m$ & The number of nodes and edges in $G$, respectively. \\ \hline
$N(v_i)$ & The set of out-neighbors of node $v_i$. \\ \hline
$d(v_i)$ & The out-degree of node $v_i$, \ie $|N(v_i)|$. \\ \hline
$c$ & The damping factor in CoSimRank. \\ \hline
$\epsilon$ & The absolute error threshold for CoSimRank values. \\ \hline
$\SM$ & The exact CoSimRank matrix (see Equation \eqref{eq:csim}). \\ \hline
$\widehat{\SM}$ & The approximate CoSimRank matrix. \\ \hline
\end{tabular}
\vspace{0mm}
\end{small}
\end{table*}

\subsection{Notations and Terminology} 
Let $G = (V, E)$ be an unweighted graph with $n$ nodes and $m$ edges, where $V$ and $E$ denote the node and edge sets, respectively. We denote by $d(v_i)$  the out-degree of node $v_i$ and by $N(v_i)$ the out-neighbors of node $v_i$. For simplicity, in the following we assume that $G$ is directed. For an undirected graph, we simply replace each undirected edge $(u, v)$ with two directed ones with opposing directions, \ie $(u, v)$ and $(v, u)$.

We denote matrices in bold uppercase, \eg $\MM$. We use $\MM[i]$ to denote the $i$-th row vector of $\MM$, and $\MM[\cdot,j]$ to denote the $j$-th column vector of $\MM$. In addition, we use $\MM[i,j]$ to denote the element at the $i$-th row and $j$-th column of $\MM$. Given an index set $\mathcal{I}$, we let $\MM[\mathcal{I}]$ (resp.\ $\MM[\cdot,\mathcal{I}]$) be the matrix block of $\MM$ that contains the row (resp.\ column) vectors of the indices in $\mathcal{I}$. Let $\AM$ be the adjacency matrix of the input graph $G$, \ie $\AM[i, j] = 1$ if  $(v_i,v_j)\in E$, otherwise $\AM[i, j] = 0$. Let $\DM$ be the diagonal out-degree matrix of $G$, \ie $\DM[i,i] = d(v_i)=\sum_{v_j\in V}{\AM[i,j]}$. We define the transition matrix (a.k.a. random walk matrix) of $G$ is defined as $\PM = \DM^{-1}\AM$. Accordingly, $\PM^{\ell}[i,j]$ signifies the probability that a $\ell$-step ($\ell \ge 1$) random walk (\ie the random walk that walks from the current node to the next node along out-going edges) from node $v_i$ would end at node $v_j$. Particularly, Lemma \ref{lem:p} proves an important property of transition matrix $\PM$.
Table \ref{tbl:notation} lists the frequently-used notations throughout the paper.

\begin{lemma}\label{lem:p}
Given any integer $k\ge 1$, $\sum_{v_j\in V}\PM^k[i,j]=1\ \forall{v_i\in V}$ holds.
\end{lemma}

\subsection{Problem Definition}
Definition \ref{def:cosimrank} presents the formal definition of CoSimRank.
\begin{definition}[CoSimRank \textnormal{\cite{rothe2014cosimrank}}]\label{def:cosimrank}
Given an input graph $G=(V,E)$ with its transition matrix $\PM$ and a damping factor $c\in (0,1)$, the CoSimRank value of nodes $v_i$ and $v_j$ ($v_i,v_j\in V$) is defined as: 
\begin{equation}\label{eq:csim}
\SM[i,j]=\sum_{\ell=0}^{\infty}{c^{\ell} \langle \PM^{\ell}[i], \PM^{\ell}[j] \rangle }.
\end{equation}
The matrix form of CoSimRank is defined as
\begin{equation}\label{eq:cmmatrix}
\SM = \sum_{\ell=0}^{\infty}{c^{\ell} \PM^{\ell}\cdot(\PM^{\ell})^{\top}}.
\end{equation}
\end{definition}

Since the computation of exact CoSimRank matrix $\SM$ is infeasible due to involving summing up an infinite series, this paper mainly focuses on $\epsilon$-approximate all-pairwise CoSimRank query, which is defined as follows.


\begin{definition}[$\epsilon$-approximate all pairwise CoSimRank query \textnormal{\cite{yu2015co}}]\label{def:apcosimrank}
Given an input graph $G=(V,E)$ and an absolute error threshold $\epsilon\in (0,\frac{1}{1-c})$, $\epsilon$-approximate all-pairwise CoSimRank query returns an $n\times n$ matrix $\widehat{\SM}$ that
\begin{equation}\label{eq:appox-cmmatrix}
\left|\widehat{\SM}[i,j]-\SM[i,j]\right|\le \epsilon,
\end{equation}
holds for every two nodes $v_i,v_j\in V$, where $\SM[i,j]$ is the exact CoSimRank value defined in Eq.~\eqref{eq:csim}.
\end{definition}

According to Lemma \ref{lem:p}, for every two nodes $v_i,v_j\in V$, we have $\SM[i,j] \le \sum_{\ell=0}^{\infty}{c^{\ell}}= \frac{1}{1-c}$. Thus, we require $0< \epsilon< \frac{1}{1-c}$.


\section{Related Work}\label{sec:rw}
In this section, we first review three algorithms for $\epsilon$-approximate all-pairwise CoSimRank query, \ie, \piter, \csmt, and \fsim, that are most related to our solutions; after that, we simply review other work related to CoSimRank computation.

\begin{table*}[!t]
\centering
\renewcommand*{\arraystretch}{2.0}
\caption{Theoretical guarantees of $\epsilon$-approximate all-pairwise CoSimRank algorithms.}\vspace{2mm}\label{tbl:algos}
\resizebox{\textwidth}{!}{
\begin{tabular}{|l|c|c|}
\hline
\bf Name & \bf Accuracy & \bf Time Complexity \\ \hline
\piter\cite{rothe2014cosimrank}     &    $|s(v_i,v_j)-\hat{s}(v_i,v_j)|\le \epsilon,\ \forall{v_i,v_j}\in V$      &      $O\left(n^3\ln(\frac{1}{\epsilon})\right)$         \\ \hline
\csmt\cite{yu2015co}     &    $|s(v_i,v_j)-\hat{s}(v_i,v_j)|\le \epsilon,\ \forall{v_i,v_j}\in V$       &       $O\left(n^3\log_2(\ln(\frac{1}{\epsilon}))\right)$               \\ \hline
\fsim\cite{yu2018fast}     &   $|s(v_i,v_j)-\hat{s}(v_i,v_j)|\le \epsilon,\ \forall{v_i,v_j}\in V$       &       $O\left(n^3\ln(\frac{1}{\epsilon})\right)$             \\ \hline
\rsim     & $\mathbb{P}\left[|s(v_i,v_j)-\hat{s}(v_i,v_j)|\le \epsilon,\ \forall{v_i,v_j}\in V\right]\ge 1-\frac{1}{n}$       &     $O\left(\min\left\{\frac{n^2\ln(n)}{\epsilon^2}\cdot \ln(\frac{1}{\epsilon}),n^3\ln(\frac{1}{\epsilon})\right\}\right)$                   \\ \hline
\end{tabular}
}
\vspace{0mm}
\end{table*}

\header
{\bf \piterb}. The \piter method is proposed in \cite{rothe2014cosimrank}. It computes a single element of $\SM$ iteratively from an inner product of two $k$-step random walk matrices, \ie $\PM^{k}$ and $\PM^{k \top}$. Specifically, \piter initializes $\widehat{\SM}^{(0)}=\IM$, and then in $k$-th iteration computes
\begin{equation}\label{eq:piter-smk}
\widehat{\SM}^{(k)}=c\PM\widehat{\SM}^{(k-1)}\PM^{\top}+\widehat{\SM}^{(0)}.
\end{equation}
Let $t=\frac{\ln((1-c)\epsilon)}{\ln(c)}-1$. After $t$ iterations, we have that for every two nodes $v_i,v_j\in V$,
\begin{align*}
\left|\widehat{\SM}^{(t)}[i,j]-\SM[i,j]\right|& = \sum_{k=0}^{\infty}{c^{k} \PM^{k}[i]\cdot\PM^{k}[j]}-\sum_{k=0}^{t}{c^{k} \PM^{k}[i]\cdot\PM^{k}[j]}\\
&=\sum_{k=t+1}^{\infty}{c^{k} \PM^{k}[i]\cdot\PM^{k}[j]}\\
&\le \sum_{k=t+1}^{\infty}{c^{k}} = \frac{c}{1-c}-\sum_{k=1}^{t}{c^k}=\frac{c^{t+1}}{1-c}=\epsilon,
\end{align*}
which exactly satisfies Equation~\eqref{eq:appox-cmmatrix}. The computation of $\widehat{\SM}$ involves a time complexity of $O(n^3\ln(\frac{1}{\epsilon}))$. This time complexity consists of two parts: the first part is for matrix multiplications in Equation~\eqref{eq:piter-smk} requires $O(n^3)$ time, while the second part comes from the $t$ iterations.

\header
{\bf \csmtb}. \csmt \cite{yu2015co} further reduces the theoretical time complexity of \piter to $O(n^3\log_2(\ln(\frac{1}{\epsilon})))$ by reorganizing Equation \eqref{eq:piter-smk} and reusing the intermediate results from previous iterations to facilitate the computation in further iterations in \piter. Initially, \csmt sets $\widehat{\SM}^{(0)}=\IM$ and $\QM^{(0)}=\PM$. Subsequently, it iteratively calculates
\begin{equation}\label{eq:csmt-matrix}
\begin{split}
&\widehat{\SM}^{(k)}= \widehat{\SM}^{(k-1)}+c^{2^k}\cdot (\QM^{(k-1)}\widehat{\SM}^{(k-1)}\QM^{(k-1)\top}),\\
&\QM^{(k)}=\QM^{(k) 2}.
\end{split}
\end{equation}
According to \cite{yu2015co}, applying Equation~\eqref{eq:csmt-matrix} with $t=\max\{0,\log_2(\frac{\ln((1-c)\epsilon)}{\ln(c)}-1)+1\}$ iterations is sufficient to produce an approximate CoSimRank matrix $\widehat{\SM}^{(t)}$ satisfying Equation~\eqref{eq:appox-cmmatrix}. Therefore, the computational time complexity of \csmt is $O(n^3\log_2(\ln(\frac{1}{\epsilon})))$.

\header
{\bf \fsimb}. \fsim \cite{yu2018fast} is based on the following ideas. First, \fsim decomposes $G$ into $G=T\oplus(G\ominus T)$, where $T$ is a “spanning polytree” and can be viewed as the old graph, while $G\ominus T$ can be viewed as the graph update. After that, due to the special “polytree” structure of $T$, the authors devised a fast algorithm to compute CoSimRank values $\widehat{\SM}_T$ over $T$. Finally, \fsim computes the changes of $\widehat{\SM}_T$ in response to the graph update $G\ominus T$. Given a set of nodes $\mathcal{I}$ and the number of iterations $t$ as inputs, \fsim returns approximate CoSimRank values $\widehat{\SM}[\cdot,\mathcal{I}]$ in $O(n^2|\mathcal{I}|t)$ time in the worst case, which leads to time complexity of $O(n^3t)$ if we let $\mathcal{I}=V$. Since \fsim also computes $\widehat{\SM}$ in an iterative way, $t$ is also required to be set as $\frac{\ln((1-c)\epsilon)}{\ln(c)}-1$ in order to attain the desired accuracy $\epsilon$. Hence, the total time complexity of \fsim for $\epsilon$-approximate all pairwise CoSimRank query is bounded by $O(n^3\ln(\frac{1}{\epsilon}))$.

In \cite{yu2018fast}, the dynamic scheme, \dsim, is proposed for CoSimRank computation over evolving graphs. Second-order CoSimRank is introduced in \cite{liao2019second} to effectively measure node similarities in social networks. Dhulipala {et al.} developed a parallel algorithm for single-source CoSimRank queries based on Graph Based Benchmark Suite (GBBS) \cite{dhulipala2020graph}.

Additionally, observe that the definition of CoSimRank (see Equation~\eqref{eq:cmmatrix}) is closely associated with PPR and SimRank. Due to the rapid advancements in approximate PPR and SimRank computations, a promising idea for the efficient CoSimRank computation might be utilizing these approximate PPR algorithms \cite{yang2020homogeneous,wang2017fora,shi2019realtime,wang2019efficient,lin2019distributed,wu2021unifying,DBLP:journals/pvldb/HouCWW21} or recent approximate SimRank algorithms \cite{shi2020realtime,wang2020exact,wei2019prsim,tian2016sling,wang2020disk}. However, most of these methods are designed for single-source PPR/SimRank queries instead of all pairwise queries and it is non-trivial to have them tailored for CoSimRank computation. Thus, in this paper, we do not discuss how to exploit these methods for CoSimRank computation and leave it as future work.

Table~\ref{tbl:algos} compares the theoretical assurance of our proposed algorithm against that of existing algorithms in terms of accuracy and complexity. Our proposed \rsim answers $\epsilon$-approximate all pairwise CoSimRank query successfully with probability $1-\frac{1}{n}$ using $O\left(\min\left\{\frac{n^2\ln(n)}{\epsilon^2}\cdot \ln(\frac{1}{\epsilon}),n^3\ln(\frac{1}{\epsilon})\right\}\right)$ time. In the following section, we elaborate details and theoretical analysis of \rsim.

\section{The \rsim Algorithm}\label{sec:algo}
This section presents our randomized algorithm, \ie \rsim, for answering $\epsilon$-approximate all-pairwise CoSimRank queries. Observe that the tremendous overheads incurred by the CoSimRank computation are caused by the $n\times n$ matrix multiplications between $\PM^{k}$ and $\PM^{k\top}$. A fundamental tool to speed up the matrix multiplication is approximating the results in a $d$-dimensional ($d\ll n$) space through random projection such that the pairwise distance can be preserved within a certain error. Johnson–Lindenstrauss transformation \cite{johnson1984extensions,achlioptas2003database,matouvsek2008variants} allows us to reduce the dimension from $n$ to $d$ that is independent of $n$. In this way, the matrix multiplications can be done in $O(n^2d)$ time. Although Johnson–Lindenstrauss transformation is mainly devised for preserving the Euclidean distance between two vectors accurately, it is also able to preserve the inner product of two vectors within a certain error, which will be shown to be sufficient for our purpose. We illustrate our proposed algorithm in Section~\ref{sec:algo}, followed by a theoretical analysis of \rsim in terms of accuracy and complexity in Section~\ref{sec:analysis}. Section \ref{sec:choose} discusses the choice of parameter $\delta$ used in \rsim to achieve the optimal running time in practice. 

\subsection{Main Algorithm}\label{sec:algo}
\begin{algorithm}[!t]
\begin{small}
\caption{\rsim}\label{alg:rsim}
\KwIn{An input graph $G$, $c, \epsilon, p_f,\delta$.}
\KwOut{$\widehat{\SM}$.}
$t\gets \left\lceil\frac{\ln(1-\frac{c-(1-c)\epsilon}{c(1-\delta)})}{\ln(c)}\right\rceil$\;
$d\gets \left\lceil\frac{2\ln(\frac{n^2}{2p_f})}{\delta-\ln(1+\delta)}\right\rceil$\;
\If{$d\ge n$}{
$\QM\gets \PM$
}\Else{
Generate $\TM\in \mathbb{R}^{n\times d}\sim \mathcal{N}(0,1)$\Comment*[r]{$O(nd)$ time}\
$\QM \gets \frac{1}{\sqrt{d}}\cdot\PM\TM$\Comment*[r]{$O(md)$ time}
}
$\HM^{(1)} \gets \sqrt{c}\cdot\QM$;\ 
$\widehat{\SM}\gets \IM+\HM^{(1)}\cdot\HM^{(1)\top}$\Comment*[r]{$O(n^2d)$ time}\
\For{$k\gets 2$ to $t$}{
    $\HM^{(k)}\gets \sqrt{c}\PM\cdot\HM^{(k-1)}$\Comment*[r]{$O(md)$ time}\
    $\widehat{\SM} \gets \widehat{\SM} + \HM^{(k)}\cdot\HM^{(k)\top}$\Comment*[r]{$O(n^2d)$ time}
}
\Return $\widehat{\SM}$\;
\end{small}
\end{algorithm}

Algorithm~\ref{alg:rsim} shows the pseudo-code of \rsim, which takes an input graph $G$, damping factor $c\in (0,1)$, absolute error threshold $\epsilon$, failure probability $p_f$, and parameter $\delta$ as inputs. Initially, \rsim calculates $t=\left\lceil\frac{\ln(1-\frac{c-(1-c)\epsilon}{c(1-\delta)})}{\ln(c)}\right\rceil$ and $d=\left\lceil\frac{2\ln(\frac{n^2}{2p_f})}{\delta-\ln(1+\delta)}\right\rceil$ (Lines 1-2). If $d\ge n$, meaning that projecting $\PM$ to an $n\times d$ matrix leads to an $O(n^3)$ time or even higher time costs, we can set $\QM=\PM$ instead to ensure that the time complexity incurred in matrix multiplications is bounded by $O(n^3)$, and thus, \rsim degrades to \piter method. Otherwise, an $n\times d$ projection matrix $\TM$ is generated, where each entry is an independent and identically distributed random variable sampled from a Gaussian $\mathcal{N}(0,1)$ (Lines 3-7), and \rsim initializes $\QM=\frac{1}{\sqrt{d}}\cdot\PM\TM$. After that, Algorithm~\ref{alg:rsim} sets $\HM^{(1)}=\QM$ and computes $\widehat{\SM}=\IM+\HM^{(1)}\cdot\HM^{(1)\top}$ (Lines 8-9). Subsequently, \rsim iteratively computes approximate CoSimRank matrix $\widehat{\SM}$ with $t-1$ iterations. Specifically, in $k$-th iteration, \rsim computes $\HM^{(k)}=\sqrt{c}\PM\cdot\HM^{(k-1)}$ and increase $\widehat{\SM}$ by $\HM^{(k)}\cdot\HM^{(k)\top}$ (Lines 10-11). Finally, Algorithm~\ref{alg:rsim} returns $\widehat{\SM}$ as an approximation of CoSimRank matrix $\SM$.

\subsection{Analysis}\label{sec:analysis}
Before analyzing the accuracy guarantee of Algorithm~\ref{alg:rsim}, we first introduce the following lemmas.
\begin{lemma}[(Preservation of inner products \cite{kaban2015improved}]\label{lem:ip}
Let $\delta, p_f\in (0,1)$ and $d\ge \frac{2\ln(1/p^{\prime}_f)}{\delta-\ln(1+\delta)}$. Let $\TM$ be an $n\times d$ matrix, where each entry is sampled i.i.d. from a Gaussian $\mathcal{N}(0,1)$. Given any two vectors $\mathbf{z}_i,\mathbf{z}_i\in \mathbb{R}^{n}$, we define $\mathbf{x}_i=\frac{1}{\sqrt{d}}\cdot \mathbf{z}_i\TM,\ \mathbf{x}_j=\frac{1}{\sqrt{d}}\cdot \mathbf{z}_j\TM$. Then, we have
\begin{align}
\mathbb{P}\left[\left|\mathbf{x}_i\cdot\mathbf{x}_j^{\top}-\mathbf{z}_i\cdot \mathbf{z}_j^{\top}\right|\le \delta\cdot ||\mathbf{z}_i||\cdot||\mathbf{z}_j||\right]\ge 1-p^{\prime}_f.
\end{align}
\end{lemma}
By applying union bound to Lemma~\ref{lem:ip}, we obtain the following corollary:
\begin{lemma}\label{lem:rp-pmatrix}
Let $\delta, p_f\in (0,1)$ and $d\ge \frac{2\ln(1/p^{\prime}_f)}{\delta-\ln(1+\delta)}$. Let $\TM$ be an $n\times d$ matrix, where each entry is sampled i.i.d. from a Gaussian $\mathcal{N}(0,1)$. We define $\QM = \frac{1}{\sqrt{d}}\cdot\PM\TM$, where $\PM$ is the transition matrix of an input graph $G$. Then, for every two nodes $v_i,v_j\in V$, the following inequality holds:
\begin{equation*}
\mathbb{P}\left[\left|\QM[i]\cdot\QM[j]^{\top}-\PM[i]\cdot \PM[j]^{\top}\right|\le \frac{\delta}{\sqrt{d(v_i)\cdot d(v_j)}}\right]\ge 1-\frac{n^2p^{\prime}_f}{2}.
\end{equation*}
\begin{proof}
First, note that for any node $v_i\in V$, 
\begin{equation}
||\PM[i]||=\sqrt{\sum_{v_j\in N(v_i)}\frac{1}{d^2(v_i)}}=\sqrt{{\frac{d(v_i)}{d^2(v_i)}}}=\sqrt{\frac{1}{d(v_i)}}.
\end{equation}
According to Lemma~\ref{lem:ip}, for any two nodes $v_i,v_j\in V$, we have
\begin{equation}
\mathbb{P}\left[\left|\QM[i]\cdot\QM[j]^{\top}-\PM[i]\cdot \PM[j]^{\top}\right|> \frac{\delta}{\sqrt{d(v_i)\cdot d(v_j)}}\right]\le p^{\prime}_f.
\end{equation}
Using union bound over all $\binom{n}{2}=\frac{n(n-1)}{2}$ node pairs, for every two nodes $v_i,v_j\in V$, we have
\begin{equation}
\mathbb{P}\left[\left|\QM[i]\cdot\QM[j]^{\top}-\PM[i]\cdot \PM[j]^{\top}\right|> \frac{\delta}{\sqrt{d(v_i)\cdot d(v_j)}}\right]\le \frac{n^2p^{\prime}_f}{2},
\end{equation}
which completes our proof.\done
\end{proof}
\end{lemma}

Based on the above analysis, we establish the accuracy guarantee of \rsim as follows:
\begin{theorem}
Given a damping factor $c\in (0,1)$, failure probability $p_f$ and an absolute error threshold $\epsilon$ as inputs to Algorithm~\ref{alg:rsim}, $\widehat{\SM}$ is returned. Then, for every two nodes $v_i,v_j\in V$,
\begin{equation*}
\left|\widehat{\SM}[i,j]-\SM[i,j]\right|\le \epsilon
\end{equation*}
holds with probability at least $1-p_f$.
\begin{proof}
By Lines 6-9, $\HM^{(k)}=\sqrt{c}^{k}\PM^{k-1}\QM$.
\begin{align*}
\widehat{\SM}-\SM&=\IM+\sum_{k=1}^{t}{\HM^{(k)} \HM^{(k)\top}}-\sum_{k=0}^{\infty}{c^{k}\PM^{k} \PM^{k\top}}\\
&=\sum_{k=1}^{t}{c^{k}\PM^{k-1}(\QM\QM^{\top})(\PM^{k-1})^{\top}}-\sum_{k=1}^{t}{c^{k}\PM^{k} \PM^{k\top}}-\sum_{k=t+1}^{\infty}{c^{k}\PM^{k} \PM^{k\top}}\\
&=\sum_{k=1}^{t}{c^{k}\PM^{k-1}(\QM\QM^{\top}-\PM\PM^{\top})(\PM^{k-1})^{\top}}-\sum_{k=t+1}^{\infty}{c^{k}\PM^{k} \PM^{k\top}}
\end{align*}
Let $\mathbf{E}$ be an $n\times n$ matrix, in which $(i,j)$ entry is equal to $\frac{\delta}{\sqrt{d(v_i)\cdot d(v_j)}}$. Using Lemma~\ref{lem:rp-pmatrix} with $d=\left\lceil\frac{2\ln(\frac{n^2}{2p_f})}{\delta-\ln(1+\delta)}\right\rceil$, we have that with probability at least $1-p_f$, we then obtain
\begin{equation*}
\|\widehat{\SM}-\SM\|_{\max} \le \left\|\sum_{k=1}^{t}{c^{k}\PM^{k-1}\mathbf{E}(\PM^{k-1})^{\top}}+\sum_{k=t+1}^{\infty}{c^{k}\PM^{k}\cdot \PM^{k\top}}\right\|_{\max}.
\end{equation*}
According to Lemma \ref{lem:p}, for every two nodes $v_i,v_j\in V$,
\begin{align*}
\|\widehat{\SM}-\SM\|_{\max} &\le \sum_{k=1}^{t}{c^{k}\delta}+\sum_{k=t+1}^{\infty}{c^{k}}= \frac{c}{1-c}-(1-\delta)\cdot\sum_{k=1}^{t}{c^{k}}\\
&=\frac{c}{1-c}(1-(1-\delta)\cdot(1-c^t)),
\end{align*}
According to Line 1 in Algorithm \ref{alg:rsim}, $t=\left\lceil\frac{\ln(1-\frac{c-(1-c)\epsilon}{c(1-\delta)})}{\ln(c)}\right\rceil$, we have $\|\widehat{\SM}-\SM\|_{\max}\le \epsilon$, \ie $\left| \widehat{\SM}[i,j]-\SM[i,j]\right|\le \epsilon\ \forall{v_i,v_j\in V}$, with probability at least $1-p_f$. The theorem is proved.\done
\end{proof}
\end{theorem}

First, the space complexity is $O(n^2)$ since we need to store the CoSimRank values for all possible node pairs in $G$. Considering the case where $d\ge n$ (Line 4 in Algorithm \ref{alg:rsim}), \rsim degrades to \piter method, and thus, the time complexity is $O\left(n^3\ln{\left(\frac{1}{\delta}\right)}\right)$. 

Next, we discuss the case where $d<n$ (Lines 6-7 in Algorithm \ref{alg:rsim}). According to \cite{achlioptas2003database}, the generation of $\TM$ requires $O(nd)$ time (Line 6). In $k$-th iteration, the computation of $\HM^{(k)}$ costs $O(md)$ time and $O(nd)$ space, while the computation of $\widehat{\SM}$ requires $O(n^2d)$ time and $O(n^2)$ space. 
By Lines 1-2 in Algorithm \ref{alg:rsim}, we need to set $d=\left\lceil\frac{2\ln(\frac{n^2}{2p_f})}{\delta-\ln(1+\delta)}\right\rceil$ and $t=\left\lceil\frac{\ln(1-\frac{c-(1-c)\epsilon}{c(1-\delta)})}{\ln(c)}\right\rceil$ iterations in total.
Therefore, the overall time complexity is 
\begin{equation}\label{eq:time}
O\left(mdt+n^2dt\right)=O\left(n^2dt\right)=O\left(n^2\ln{\left(\frac{n^2}{2p_f}\right)}\cdot \frac{\ln{\left(\frac{c(1-\delta)}{(1-c)\epsilon-c\delta}\right)}}{\delta-\ln{(1+\delta)}}\right),
\end{equation}
Since $\delta\in (0,1)$ and $(1-c)\epsilon-c\delta>0$, we obtain $0<\delta < \frac{(1-c)\cdot \epsilon}{c}$. According to the inequality $\ln{(1+\delta)} \le \delta - \delta^2/2 + \delta^3/3\ \forall{\delta} \ge 0$, the time complexity in Equation \eqref{eq:time} is bounded by $O(\frac{n^2\ln(n)}{\epsilon^2}\ln(\frac{1}{\epsilon}))$ when letting $\delta=\frac{(1-c)\cdot\epsilon}{2c}$. 
To sum up, the time complexity of \rsim is $$O\left(\min\left\{\frac{n^2\ln(n)}{\epsilon^2}\cdot \ln(\frac{1}{\epsilon}),n^3\ln(\frac{1}{\epsilon})\right\}\right).$$ 
In the subsequent subsection, we elaborate how to minimize the practical time cost in Equation \eqref{eq:time} by carefully picking $\delta$.

\header
{\bf Remark.} According to \cite{chen2015johnson,venkatasubramanian2011johnson}, setting $d$ to $\left\lceil\frac{\ln(\frac{n^2}{2p_f})}{2(\delta-\ln(1+\delta))}\right\rceil$ is good enough for Johnson–Lindenstrauss transformation to achieve the desired accuracy in practice. The gap between the theoretic bound and practical one is due to the union bound used in the proof of Lemma~\ref{lem:rp-pmatrix}.

\begin{algorithm}[!t]
\begin{small}
\caption{$\mathsf{TernarySearch}$}\label{alg:eps}
\KwIn{$c, \epsilon$.}
\KwOut{$\delta$.}
$\delta_{l}\gets 0,\ \delta_{u}\gets \frac{1-c}{c}\cdot\epsilon$\;
\While{{\bf true}}{
$\delta^{\prime}_l\gets \delta_l+\frac{\delta_u-\delta_l}{3}$\;
$\delta^{\prime}_u\gets \delta_u-\frac{\delta_u-\delta_l}{3}$\;
\lIf{$\delta^{\prime}_u\le \delta^{\prime}_l$ or $\delta_u-\delta_l\le \frac{1-c}{1000c}\cdot\epsilon$}{{\bf break}}
\If{$f(\delta^{\prime}_l)<f(\delta^{\prime}_u)$}{
$\delta_{l}\gets \delta^{\prime}_l$\;
}
\Else{
$\delta_{u}\gets \delta^{\prime}_u$\;
}
}
$\delta \gets \frac{\delta_{l}+\delta_{u}}{2}$\;
\Return $\delta$\;
\end{small}
\end{algorithm}

\subsection{Choosing $\delta$}\label{sec:choose}
In the following, we explain the rationale of Algorithm~\ref{alg:eps} for choosing $\delta$. According to Equation \eqref{eq:time}, the total time complexity of \rsim is linear to 
\begin{equation}\label{eq:feps}
f(\delta)=\frac{\ln{\left(\frac{c(1-\delta)}{(1-c)\epsilon-c\delta}\right)}}{\delta-\ln{(1+\delta)}}. 
\end{equation}
Recall that $0<\delta < \frac{(1-c)}{c}\cdot \epsilon$. Therefore, to achieve the minimum time cost, the aim is then to find a value for $\delta$ such that $f(\delta)$ is minimized, \ie
\begin{equation}
\min_{0<\delta<\frac{(1-c)}{c}\cdot \epsilon}{f(\delta)}.
\end{equation}
$f(\delta)$ is a convex function when $\delta$ is in range $(0, \frac{(1-c)}{c}\cdot \epsilon)$. Thus, we can find an approximate minimum value for $f(\delta)$ by using a ternary search algorithm \cite{levitin2012introduction}, as presented in Algorithm~\ref{alg:eps}. 
Specifically, Algorithm~\ref{alg:eps} takes damping factor $c$ and error threshold $\epsilon$ as inputs and then computes the initial lower bound $\delta_l=0$ and upper bound $\delta_u=\frac{1-c}{c}\cdot\epsilon$ for $\delta$ (Line 1). Subsequently, Algorithm~\ref{alg:eps} starts iterations to update the lower bound $\delta_l$ and upper bound $\delta_u$. In each iteration, Algorithm~\ref{alg:eps} first computes $\delta^{\prime}_l= \delta_l+\frac{\delta_u-\delta_l}{3}$ and $\delta^{\prime}_u= \delta_u-\frac{\delta_u-\delta_l}{3}$ (Lines 3-4). If $\delta^{\prime}_u\le\delta^{\prime}_l$ or $\delta_u-\delta_l\le \frac{1-c}{1000c}\cdot\epsilon$, the lower bound $\delta_l$ is very close to the upper bound $\delta_u$, and thus, we can terminate the search of the lower and upper bounds (Line 5). When the termination condition is not satisfied, Algorithm~\ref{alg:eps} updates the lower bound $\delta_l=\delta^{\prime}_l$ if  $f(\delta^{\prime}_l)<f(\delta^{\prime}_u)$, otherwise $\delta_u=\delta^{\prime}_u$ (Lines 6-9), and proceeds to next iteration. Finally, Algorithm~\ref{alg:eps} computes $\delta=\frac{\delta_{l}+\delta_{u}}{2}$ and returns it (Lines 10-11).
Note that when $\delta_u-\delta_l\le \frac{1-c}{1000c}\cdot\epsilon$, Algorithm~\ref{alg:eps} finishes searching. Hence, Algorithm~\ref{alg:eps} requires $\log_3{\left(\frac{(1-c)\cdot\epsilon}{c}\cdot \frac{1000c}{(1-c)\cdot \epsilon}\right)}=\log_3{(1000)}$ iterations in a worst case.

\section{Experiments}\label{sec:exp}
We experimentally evaluate our proposed \rsim against two competitors in terms of efficiency on 6 real datasets. All experiments are conducted on a Linux machine powered by an Intel Xeon(R) Gold 6240@2.60GHz CPU and 377GB RAM. Source codes of all methods are implemented in Python and all matrices are represented in a sparse form to avoid unnecessary space overheads.

\begin{table*}[!t]
\centering
\renewcommand*{\arraystretch}{1.4}
\begin{small}
\caption{Statistics for Datasets.}\vspace{2mm}\label{tbl:data}
\begin{tabular}{|l|c|c|c|}
\hline
\bf Name & \bf \#Nodes ($n$) & \bf \#Edges ($m$) & \bf Type\\ \hline
{\em Facebook} & 4,039 & 88,234 & undirected \\ \hline
{\em as-735}  & 7,716 & 26,467 & undirected   \\ \hline
{\em ca-HepPh} & 12,008  & 237,010 & undirected \\ \hline
{\em email-Enron} & 36,692 & 183,831 & directed \\ \hline
{\em Twitter} &  81,306  & 1,768,149  & directed \\ \hline
{\em Google+} & 107,614 &	13,673,453 & directed \\ \hline
\end{tabular}
\vspace{0mm}
\end{small}
\end{table*}

\subsection{Experimental Setting}
\header
{\bf Datasets.} We experiment with six real-world graphs that are used in previous work \cite{yu2015co,yu2018fast}, which are taken from \cite{snapnets}. Table \ref{tbl:data} lists the statistics of the datasets. {\em as-735 (AS)}\footnote{\url{http://www.cise.ufl.edu/research/sparse/matrices/SNAP/as-735.html}} is a communication network of autonomous systems extracted from the Border Gateway Protocol logs, where an edge represents a who-talks-to-whom relationship. {\em ca-HepPh}\footnote{\url{http://snap.stanford.edu/data/ca-HepPh.html}} is a collaboration graph from the arXiv High Energy Physics, where each node is an author and each edge represents a collaboration relationship. {\em email-Enron}\footnote{\url{http://snap.stanford.edu/data/email-Enron.html}} is an email communication network collected from Enron, in which each node signifies an email address and there is an edge between two nodes if at least one email is sent between them. {\em Facebook}\footnote{\url{http://snap.stanford.edu/data/ego-Facebook.html}}, {\em Twitter}\footnote{\url{http://snap.stanford.edu/data/ego-Twitter.html}}, and {\em Google+}\footnote{\url{http://snap.stanford.edu/data/ego-Gplus.html}} are social networks used in \cite{mcauley2012learning}. 

\header
{\bf Parameter Settings.} We compare \rsim against \piter, \csmt, and \fsim in terms of efficiency for $\epsilon$-approximate all-pairwise CoSimRank queries. Following prior work \cite{yu2015co}, we set damping factor $c=0.8$, meaning that $\epsilon$ should be in range $(0, 5)$. In our experiments, we vary absolute error threshold $\epsilon$ in range $\{1.0, 0.5, 0.2, 0.1, 0.05\}$. We report
the running time (measured in wall-clock time) of each algorithm on each dataset with various $\epsilon$ settings. Note that the $y$-axis is in log-scale and the measurement unit for running time is second (sec). We omit any methods if they can not terminate within two days or run out of memory.

\begin{figure*}[!t]
\centering
\begin{small}
\begin{tikzpicture}
    \begin{customlegend}[legend columns=4,
        legend entries={\rsim,\piter,\csmt,\fsim},
        legend style={at={(0.45,1.05)},anchor=north,draw=none,font=\footnotesize,column sep=0.2cm}]
    \addlegendimage{line width=0.25mm,mark=o}
    \addlegendimage{line width=0.25mm,mark=diamond}
    \addlegendimage{line width=0.25mm,mark=square}
    \end{customlegend}
\end{tikzpicture}
\\[-\lineskip]
\vspace{0mm}
\subfloat[{\em Facebook}]{
\begin{tikzpicture}[scale=1]
    \begin{axis}[
        height=\columnwidth/3.3,
        width=\columnwidth/2.6,
        ylabel={\em running time (sec)},
        xmin=0.5, xmax=5.5,
        ymin=1, ymax=2000,
        xtick={1,2,3,4,5,6},
        xticklabel style = {font=\tiny},
        yticklabel style = {font=\scriptsize},
        xticklabels={1.0,0.5,0.2,0.1,0.05},
        scaled y ticks = false,
        ymode=log,
        log basis y={10},
        every axis y label/.style={at={(current axis.north west)},right=13mm,above=0mm},
        every axis x label/.style={at={(current axis.right of origin)},above=10mm,anchor=north west},
    ]
    \addplot[line width=0.25mm,mark=o]  
        plot coordinates {
(1,	2.713598013	)
(2,	7.580753088	)
(3,	20.79153990	)
(4,	56.27616191	)
(5,	70.38353801	)
    };
    
    \addplot[line width=0.25mm,mark=diamond]  
        plot coordinates {
(1,	16.8331511	)
(2,	28.83103395	)
(3,	44.96079397	)
(4,	56.27616191	)
(5,	70.38353801	)
    };

    \addplot[line width=0.25mm,mark=square]  
        plot coordinates {
(1,	99.3682045	)
(2,	172.9402909	)
(3,	211.1574346	)
(4,	292.8601711	)
(5,	479.0551928	)
    };
    
    \addplot[line width=0.25mm,mark=o]  
        plot coordinates {

    };
    \end{axis}
\end{tikzpicture}%
}%
\subfloat[{\em as-735}]{
\begin{tikzpicture}[scale=1]
    \begin{axis}[
        height=\columnwidth/3.3,
        width=\columnwidth/2.6,
        ylabel={\em running time (sec)},
        xmin=0.5, xmax=5.5,
        ymin=1, ymax=3000,
        xtick={1,2,3,4,5,6},
        xticklabel style = {font=\tiny},
        yticklabel style = {font=\scriptsize},
        xticklabels={1.0,0.5,0.2,0.1,0.05},
        scaled y ticks = false,
        ymode=log,
        log basis y={10},
        every axis y label/.style={at={(current axis.north west)},right=13mm,above=0mm},
        every axis x label/.style={at={(current axis.right of origin)},above=10mm,anchor=north west},
    ]
    \addplot[line width=0.25mm,mark=o]  
        plot coordinates {
(1,	4.357288837	)
(2,	9.436233044	)
(3,	24.75606393	)
(4,	47.20589399	)
(5,	78.35401917	)
    };
    
    \addplot[line width=0.25mm,mark=diamond]  
        plot coordinates {
(1,	20.4177599	)
(2,	29.57783794	)
(3,	47.62933278	)
(4,	55.86095905	)
(5,	78.35401917	)
    };

    \addplot[line width=0.25mm,mark=square]  
        plot coordinates {
(1,	227.6618156	)
(2,	398.2997899	)
(3,	417.3799953	)
(4,	493.4434669	)
(5,	569.1545391	)
    };
    
    \addplot[line width=0.25mm,mark=o]  
        plot coordinates {

    };
    \end{axis}
\end{tikzpicture}%
}%
\subfloat[{\em ca-HepPh}]{
\begin{tikzpicture}[scale=1]
    \begin{axis}[
        height=\columnwidth/3.3,
        width=\columnwidth/2.6,
        ylabel={\em running time (sec)},
        xmin=0.5, xmax=5.5,
        ymin=10, ymax=5000,
        xtick={1,2,3,4,5,6},
        xticklabel style = {font=\tiny},
        yticklabel style = {font=\scriptsize},
        xticklabels={1.0,0.5,0.2,0.1,0.05},
        scaled y ticks = false,
        ymode=log,
        log basis y={10},
        every axis y label/.style={at={(current axis.north west)},right=13mm,above=0mm},
        every axis x label/.style={at={(current axis.right of origin)},above=10mm,anchor=north west},
    ]
    \addplot[line width=0.25mm,mark=o]  
        plot coordinates {
(1,	14.10146999	)
(2,	27.62072706	)
(3,	98.26852893	)
(4,	201.6313076	)
(5,	396.1222789	)
    };

    \addplot[line width=0.25mm,mark=diamond]  
        plot coordinates {
(1,	126.6734409	)
(2,	153.0408089	)
(3,	248.0157771	)
(4,	306.1054771	)
(5,	396.1222789	)
    };

    \addplot[line width=0.25mm,mark=square]  
        plot coordinates {
(1,	520.8672309	)
(2,	989.1538832	)
(3,	1302.168077	)
(4,	1736.224103	)
(5,	2282.662807	)
    };
    \end{axis}
\end{tikzpicture}%
}%

\subfloat[{\em email-Enron}]{
\begin{tikzpicture}[scale=1]
    \begin{axis}[
        height=\columnwidth/3.3,
        width=\columnwidth/2.6,
        ylabel={\em running time (sec)},
        xmin=0.5, xmax=5.5,
        ymin=100, ymax=20000,
        xtick={1,2,3,4,5,6},
        xticklabel style = {font=\tiny},
        yticklabel style = {font=\scriptsize},
        xticklabels={1.0,0.5,0.2,0.1,0.05},
        scaled y ticks = false,
        ymode=log,
        log basis y={10},
        every axis y label/.style={at={(current axis.north west)},right=13mm,above=0mm},
        every axis x label/.style={at={(current axis.right of origin)},above=10mm,anchor=north west},
    ]
    \addplot[line width=0.25mm,mark=o]  
        plot coordinates {
(1,	171.3863289	)
(2,	298.1799939	)
(3,	566.2160029	)
(4,	1297.707438	)
(5,	2513.030325	)
    };

    \addplot[line width=0.25mm,mark=diamond]  
        plot coordinates {
(1,	647.8417702	)
(2,	1049.806121	)
(3,	1637.784221	)
(4,	2054.854179	)
(5,	2513.030325	)
    };

    \addplot[line width=0.25mm,mark=square]  
        plot coordinates {
(1,	2663.853973	)
(2,	6785.247729	)
(3,	8598.930096	)
(4,	11655.09153	)
(5,	14481.38911	)
    };
    \end{axis}
\end{tikzpicture}%
}%
\subfloat[{\em Twitter}]{
\begin{tikzpicture}[scale=1]
    \begin{axis}[
        height=\columnwidth/3.3,
        width=\columnwidth/2.6,
        ylabel={\em running time (sec)},
        xmin=0.5, xmax=5.5,
        ymin=500, ymax=200000,
        xtick={1,2,3,4,5,6},
        xticklabel style = {font=\tiny},
        yticklabel style = {font=\scriptsize},
        xticklabels={1.0,0.5,0.2,0.1,0.05},
        scaled y ticks = false,
        ymode=log,
        log basis y={10},
        every axis y label/.style={at={(current axis.north west)},right=13mm,above=0mm},
        every axis x label/.style={at={(current axis.right of origin)},above=10mm,anchor=north west},
    ]
    \addplot[line width=0.25mm,mark=o]  
        plot coordinates {
(1,	869.6374012	)
(2,	1664.517385	)
(3,	4542.851165	)
(4,	13109.46461	)
(5,	55846.31923	)
    };

    \addplot[line width=0.25mm,mark=diamond]  
        plot coordinates {
(1,	26382.1087	)
(2,	44071.4499	)
(3,	81079.76669	)
(4,	104156.3157	)
(5,	134785.6873	)
    };

    \addplot[line width=0.25mm,mark=square]  
        plot coordinates {

    };
    \end{axis}
\end{tikzpicture}%
}%
\subfloat[{\em Google+}]{
\begin{tikzpicture}[scale=1]
    \begin{axis}[
        height=\columnwidth/3.3,
        width=\columnwidth/2.6,
        ylabel={\em running time (sec)},
        xmin=0.5, xmax=4.5,
        ymin=1000, ymax=100000,
        xtick={1,2,3,4,5},
        xticklabel style = {font=\tiny},
        yticklabel style = {font=\scriptsize},
        xticklabels={1.0,0.5,0.2,0.1},
        scaled y ticks = false,
        ymode=log,
        log basis y={10},
        every axis y label/.style={at={(current axis.north west)},right=13mm,above=0mm},
        every axis x label/.style={at={(current axis.right of origin)},above=10mm,anchor=north west},
    ]
    \addplot[line width=0.25mm,mark=o]  
        plot coordinates {
(1,	1796.647597	)
(2,	3927.800015	)
(3,	12833.95033	)
(4,	44442.4556	)
    };

    \addplot[line width=0.25mm,mark=diamond]  
        plot coordinates {

    };

    \addplot[line width=0.25mm,mark=square]  
        plot coordinates {

    };
    \end{axis}
\end{tikzpicture}%
}%
\vspace{-3mm}
\end{small}
\caption{Running time with varying $\epsilon$.} \label{fig:time}
\vspace{-3mm}
\end{figure*}

\subsection{Efficiency Evaluation}
Figure \ref{fig:time} plots the running time of \rsim, \piter and \csmt on six datasets when varying $\epsilon$ in $\{1.0, 0.5, 0.2, 0.1, 0.05\}$. First, observe that \csmt runs much slower than \piter, which is inconsistent with their theoretical time complexities as introduced in Section \ref{sec:rw}. The reason is that the transition matrix $\PM$ is often very sparse in practice, making the empirical running time of \piter far less than its theoretical time. Moreover, \piter only requires updating $\widehat{\SM}^{k}$ based on Equation \eqref{eq:piter-smk}, while \csmt needs to compute $\widehat{\SM}^{k}$ and $\QM^{(k)}$ in each iteration, which involves 
additional matrix multiplications. Another observation we can make from Figure \ref{fig:time} is that, on small graphs including {\em Facebook}, {\em as-735} and {\em ca-HepPh}, \rsim is $2$-$9\times$ faster than \piter when $\epsilon\ge 0.2$, and requires almost the same time as \piter when $\epsilon\le 0.1$. This is due to that on small graphs, a small $\epsilon$ value is likely to lead to $d\ge n$, and thus, \rsim degrades to \piter. On the million-edge graph {\em Twitter}, \rsim consistently outperforms \piter by up to three orders of magnitude. For instance, when $\epsilon=0.1$, \rsim requires 3.7 hours while \piter costs about 1.2 days on {\em Twitter} dataset. For the largest {\em Google+} dataset, our solution \rsim is the only viable solution to obtain approximate all pairwise CoSimRank values when varying $\epsilon$ from $0.1$ to $1.0$ on a single server, while both \piter and \csmt run out of memory and fail to return results. This implies that \rsim consumes much fewer space costs in practice compared with \piter and \csmt.
\section{Conclusion}\label{sec:clc}
This paper presents \rsim, a random projection-based method for answering $\epsilon$-approximate all pairwise CoSimRank query with $\epsilon$ worst-case absolute error in each CoSimRank value with a high probability. \rsim requires $O(\frac{n^2\ln(n)}{\epsilon^2}\ln(\frac{1}{\epsilon}))$ time to process all node pairs in the graph. In many practical scenarios, a relatively large absolute error guarantee $\epsilon$ in each CoSimRank value is sufficient for our purpose to identify similar nodes. \rsim is up to orders of magnitude faster than prior work in such scenarios. However, \rsim suffers from an expensive time complexity of $O(n^4)$ when high-precision results are desired, \eg $\frac{1}{n}$. In the future work, we will study how to answer $\epsilon$-approximate all pairwise CoSimRank query in $O(n^2\ln{(n)}\ln{(\frac{1}{\epsilon})})$ time.
%
%
%
\bibliographystyle{splncs04}
\bibliography{main}

\end{document}